\documentclass{article}

\usepackage{arxiv}

\usepackage[utf8]{inputenc} 
\usepackage[T1]{fontenc}    
\usepackage{hyperref}       
\usepackage{url}            
\usepackage{booktabs}       
\usepackage{amsfonts}       
\usepackage{nicefrac}       
\usepackage{microtype}      
\usepackage{lipsum}		
\usepackage{graphicx}
\usepackage{natbib}
\usepackage{doi}

\title{Compressive Sensing Imaging Using Caustic Lens Mask Generated by Periodic Perturbation in a Ripple Tank}

\author{ \href{https://orcid.org/0000-0002-2636-3016}{\includegraphics[scale=0.06]{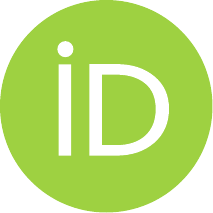}\hspace{1mm}Doğan Tunca~Arık}\thanks{Corresponding author} \\
	Electrical and Electronics Engineering Department\\
	Ankara Yıldırım Beyazıt University\\
	Ankara, Turkey \\
	\texttt{205105404@aybu.edu.tr} \\
	\And
	\href{https://orcid.org/0000-0001-9759-8448}{\includegraphics[scale=0.06]{orcid.pdf}\hspace{1mm}Asaf Behzat~Şahin} \\
	Electrical and Electronics Engineering Department\\
	Ankara Yıldırım Beyazıt University\\
	Ankara, Turkey\\
	\texttt{absahin@aybu.edu.tr} \\
        \And
	\href{https://orcid.org/0000-0002-0160-1161}{\includegraphics[scale=0.06]{orcid.pdf}\hspace{1mm}Özgün~Ersoy} \\
	Electrical and Electronics Engineering Department\\
	Ankara Yıldırım Beyazıt University\\
	Ankara, Turkey\\
	\texttt{ozgun.ersoy@aybu.edu.tr} \\}

\date{}

\renewcommand{\shorttitle}{\textit{} }

\hypersetup{
pdftitle={paper},
pdfsubject={eess.SP},
pdfauthor={Doğan T.~Arık, Asaf B.~Şahin, Özgün~Ersoy},
pdfkeywords={Compressive Sensing, Caustic Lens, Convolutional Neural Network, Random Sampling Mask},
}

\begin{document}
\maketitle

\begin{abstract}
     Terahertz imaging shows significant potential across diverse fields, yet the cost-effectiveness of multi-pixel imaging equipment remains an obstacle for many researchers. To tackle this issue, the utilization of single-pixel imaging arises as a lower-cost option, however, the data collection process necessary for reconstructing images is time-consuming. Compressive Sensing offers a promising solution by enabling image generation with fewer measurements than required by Nyquist's theorem, yet long processing times remain an issue, especially for large-sized images. Our proposed solution to this issue involves using caustic lens effect induced by perturbations in a ripple tank as a sampling mask. The dynamic characteristics of the ripple tank introduce randomness into the sampling process, thereby reducing measurement time through exploitation of the inherent sparsity of THz band signals. In this study, a Convolutional Neural Network was used to conduct target classification, based on the distinctive signal patterns obtained via the caustic lens mask. The suggested classifier obtained a 95.16 \% accuracy rate in differentiating targets resembling Latin letters.
\end{abstract}

\keywords{Compressive Sensing \and Caustic Lens \and Convolutional Neural Network \and Random Sampling Mask}

\section{Introduction}
\label{sec:Introduction}
Terahertz (THz) imaging has emerged as a highly attractive field of research with significant interest across diverse scientific disciplines [\cite{strkag2023non, afsah2019comprehensive, valuvsis2021roadmap, taylor2011thz, stantchev2020real}]. Despite the immense potential of Terahertz (THz) imaging, the high cost of traditional multi-pixel setups poses a significant barrier for many users. Single-pixel imaging configurations present a promising solution, offering a more cost-effective approach to address this challenge [\cite{stantchev2020real}]. However, the process of collecting data necessary for image reconstruction is time-consuming because it requires the number of measurements to be at least equal to the number of pixels in the image [\cite{zanotto2020single,hu2022advances,yang2020terahertz}] Over the years, various methodologies have been devised to improve the efficiency of image acquisition in the THz band. The Compressive Sensing (CS) technique, which enables the generation of images using fewer samples than Nyquist's theorem requires, presents a potentially important solution [\cite{baraniuk2007compressive,candes2008introduction}]

\begin{figure}
    \centering
    \includegraphics[width=0.5\linewidth]{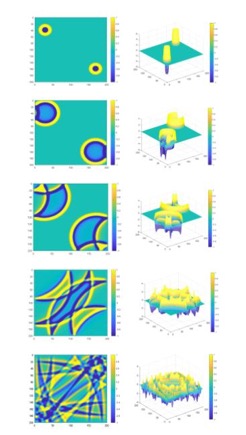}
    \caption{A mechanically controlled pump creates intricate caustic patterns on the ripple tank’s surface}
    \label{fig:fig1}
\end{figure}

The initial development of single-pixel THz imaging setup based on CS was pioneered by Chan et al. in 2008. The suggested setup described in reference [\cite{chan2008terahertz}] can reconstruct THz images with 4096 pixels using only 500 measurements. Through the utilization of a series of binary metal masks and the reconstruction of THz images with 300 measurements, Chan et al. [\cite{chan2008single}] further improved upon this approach. Although significant progress has been made in employing CS within the single-pixel imaging setup, it still faces challenges related to long collection and reconstruction time due to the extensive number of measurements, especially for larger images. Our proposed approach to address this issue involves using the caustic lens effect produced by perturbations in a ripple tank as a sampling mask. A mechanically controlled pump induces intricate caustic patterns on the surface of a ripple tank, as depicted in Figure~\ref{fig:fig1} These patterns serve as a caustic lens mask, introducing randomness into the sampling process, thereby significantly reducing measurement time by exploiting the inherent sparsity of THz band signals. 

Distinct signal patterns have been observed for different targets in our imaging setup upon recording the signals after they pass the caustic lens mask. In this study, the Convolutional Neural Network (CNN) has been applied to execute the classification task by using the extracted features derived from these distinct signal patterns. While CNNs have primarily been employed in tasks involving two-dimensional images [\cite{scott2017training, maggiori2016convolutional}], they can be adapted for use with one-dimensional signals like time series or sequential data through appropriate adjustments [\cite{huang2019ecg, jun2018ecg}]. This research utilizes the Continuous Wavelet Transform (CWT), a frequently employed technique for converting one-dimensional signals into two-dimensional images that are suitable for integration into a CNN [\cite{he2018automatic, li2019ventricular, zhao2019integrating}]. The findings demonstrate that our classifier attained a classification accuracy of 91.03 \% for identifying targets represented as Latin letters. The integration of the caustic lens mask into the imaging setup significantly contributed to achieving this high accuracy

\section{Methodology}
\label{sec:Methodology}

Researchers have traditionally investigate waves using an instrument called a ripple tank [\cite{kuwabara1986water}]. A ripple tank generates waves when a vibrating object disrupts the surface of the water. These waves subsequently propagate outward, creating ripple patterns that are observable. Observing ripples generated in a ripple tank can provide qualitative insights into the refraction of light as it traverses through water. The peaks, or crests, of waves represent regions where the water surface is elevated. When light interacts with these elevated regions, it undergoes refraction, bending toward the normal. This bending effect resembles the converging effect of a convex lens in optics. Conversely, the troughs of waves denote areas where the water surface is lowered. Light interacting with these regions also undergoes refraction, but in the opposite direction, bending away from the normal. This is similar to the diverging effect of a concave lens in optics.

A caustic lens refers to the optical phenomenon arising from the refraction or reflection of light by a surface with varying curvature or refractive properties. This phenomenon results in the formation of concentrated patterns of light intensity, known as caustics. The dynamic ripples in the tank function as a caustic lens, influencing the trajectory and characteristics of THz waves as they propagate across the oil surface. Compressive Sensing (CS) with random sampling masks finds applications in scenarios where acquiring a full set of measurements is impractical or costly. In this study, the caustic lens effect induced by a mechanical arm is utilized as a random sampling matrix.

\begin{figure}
    \centering
    \includegraphics[width=0.25\linewidth]{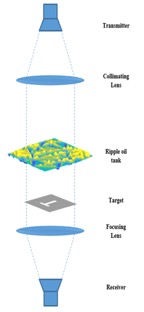}
    \caption{Proposed CS Imaging Setup}
    \label{fig:fig2}
\end{figure}

The inherent sparsity of signals in the THz bands enables the utilization of solutions based on Compressive Sensing [\cite{sarieddeen2021overview}]. Proposed CS imaging setup is shown in Figure~\ref{fig:fig2}.

The selection of the random sampling mask plays a crucial role in the performance of Compressive Sensing (CS). The non-uniformity introduced by the caustic lens mask enhances the sparsity of signals reaching the receiver of the setup. Signal sparsity is a fundamental concept in CS, indicating that only specific regions of the signal actively contribute to the measurements. This allows CS to concentrate on the essential, non-zero components of the signal.

Upon obtaining signals from the receiver of the setup, they undergo transformation into scalograms through the Continuous Wavelet Transform. Subsequently, RGB image generation from scalograms is facilitated by mapping the CWT coefficients onto color channels. These resultant images are then utilized as input for a Convolutional Neural Network (CNN) tasked with the classification of targets of various shapes.

\section{Results}
\label{sec:Results}

In this study, the 5-fold cross-validation technique was utilized to assess the classification performance. The dataset was divided into five folds, with one fold designated for testing the network and the remaining four folds for training. This process was iterated five times to obtain a less biased estimation of the model's accuracy compared to using a single data split. 

\begin{figure}
    \centering
    \includegraphics[width=0.75\linewidth]{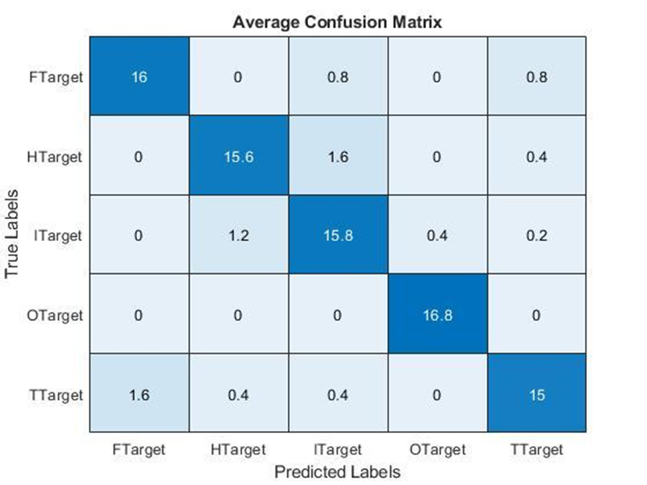}
    \caption{Average Confusion Matrix}
    \label{fig:fig3}
\end{figure}

At the end of each iteration, a confusion matrix was generated, providing a numerical representation of the predicted classes versus the actual classes in the test set. Five different confusion matrices were obtained, one for each fold. To present the overall classification performance, the average of the confusion matrices from each fold was calculated. This approach yields a more robust estimate of the classifier's performance and mitigates the impact of random variations that may occur in a single fold. Figure~\ref{fig:fig3} illustrates the average confusion matrix obtained for the network.

To evaluate the classification performance for each label, the accuracy, recall, precision, and F-measure metrics have been calculated by using the average confusion matrix present in Table~\ref{tab:table} 

These metrics offer insights into the model's performance by indicating its ability to correctly predict instances belonging to each label, capturing the true positive rate, precision of positive predictions, and a balanced measure of precision and recall, respectively.

Using the average confusion matrix for computing performance metrics ensures a more robust and representative assessment, accommodating variations within individual folds. The calculated accuracy, recall, precision, and F-measure offer valuable insights into the overall performance and effectiveness of our classification.

Our CNN classifier achieved an accuracy of 97.66\% in correctly classifying targets in the H shape, with a misclassification rate of approximately 2.34\% as F-shaped targets. For targets with a T shape, the classifier achieved a 96.09\% accurate classification, misclassifying approximately 2.34\% as I-shaped targets and 1.34\% as O-shaped targets. Achieving 100\% accuracy for O-shaped targets underscores the robust performance of the classifier in this category. While the classifier exhibited a success rate of 96.09\% in classifying I-shaped targets, it occasionally misclassified some as O and F shapes. The lowest classification accuracy, at 85.94\%, has been observed in the classification of F-shaped targets.

\begin{table}
	\caption{Performance Metric Results for Each Label}
	\centering
	\begin{tabular}{lllll}
		\toprule
		Label     & Recall     & Precision     &F-measure     &Accuracy  \\
		\midrule
		F Target & 0,9091 & 0,9091 & 0,9091 & 0,9091 \\
		H Target & 0,8864 & 0,9070 & 0,8966 & 0,8864 \\
		I Target & 0,8977 & 0,8495 & 0,8729 & 0,8977 \\
            O Target & 1,0000 & 0.9767 & 0.9882 & 1,0000 \\
            T Target & 0,8621 & 0,9146 & 0,8876 & 0,8621 \\
		\bottomrule
	\end{tabular}
	\label{tab:table}
\end{table}

\section{Conclusion}
\label{sec:Conclusion}

Based on our literature review, it is evident that the single-pixel THz imaging configuration still encounters challenges associated with long processing times. To address this issue, we propose leveraging the caustic patterns generated by perturbations in the ripple tank, thereby creating a caustic lens effect within the CS imaging setup. This caustic lens effect introduces randomness into the sampling process, thereby significantly reducing measurement time by exploiting the inherent sparsity of signals in the THz band.

Data acquired from the receiver side has been transformed into images using the colorized Continuous Wavelet Transform (CWT) technique. Subsequently, these images have been inputted into a Convolutional Neural Network (CNN) for the classification of targets with different shapes. This methodology, common in signal processing applications, involves analyzing one-dimensional signals, such as time series data, using image processing techniques. Our approach has yielded an overall classification accuracy of 95.16\% for targets of varying shapes.

While our classifier demonstrates high overall classification accuracy, as evidenced by the confusion matrix, it struggles to adequately classify targets in the form of F and T letters. By incorporating additional hidden layers into the CNN architecture, we can potentially enhance the classification process within the CS imaging setup. However, it is important to note that such improvements may come at the expense of increased processing time, thereby compromising the advantages gained in terms of efficiency.
\vspace{5mm} 

\textbf{Funding} This work was supported in part by the Scientific and Technological Research Council of Turkey (TUBITAK) under Grant 122E409.

\bibliographystyle{unsrtnat}
\bibliography{references}  






\end{document}